\documentclass[11pt]{article}
\usepackage{graphicx} 
\usepackage{listings}
\usepackage{enumitem}
\usepackage{geometry}
\usepackage{multirow}
\usepackage{graphicx}
\usepackage{pdflscape}
\usepackage{moresize}
\usepackage[colorlinks=true, allcolors=blue]
{hyperref}
\usepackage[utf8]{inputenc}
\usepackage{natbib}
\usepackage{amsmath}
\usepackage{algorithm}
\usepackage{algpseudocode}
\usepackage{array}
\usepackage{changepage}
\usepackage[english]{babel}
\usepackage{amsthm}
\usepackage{longtable} 
\usepackage{booktabs} 
\usepackage{tabularx}

\newtheorem{remark}{Remark}

\begin{document}

\title{Skill Dominance Analysis of \\ Two(Four)-player, Three(Five)-dice Variant of the Ludo Game}
\author{Tathagata Banerjee$^1$ \& Diganta Mukherjee$^2$\\
$^1$Indian Institute of Technology, Kanpur \
$^2$Indian Statistical Institute, Kolkata}
\maketitle

\begin{abstract}
    This paper examines two different variants of the Ludo game, involving multiple dice and a fixed number of total turns. Within each variant, multiple game lengths (total no. of turns) are considered.
To compare the two variants, a set of intuitive, rule-based strategies is designed, representing different broad methods of strategic play. Game play is simulated between bots (automated software applications executing repetitive tasks over a network) following these strategies.
The expected results are computed using certain game theoretic and probabilistic explanations, helping to understand the performance of the different strategies.

The different strategies are  further analyzed using win percentage in a large number of simulations, and Nash Equilibrium strategies are computed for both variants for a varying number of total turns. The Nash Equilibrium strategies across different game lengths are compared. 
A clear distinction between performances of strategies is observed, with more sophisticated strategies beating the naive one. A gradual shift in optimal strategy profiles is observed with changing game length, and certain sophisticated strategies even confound each other's performance while playing against each other.
\end{abstract}

\section{Introduction}

In this world, games serve as both entertainment and cognitive exercises, fostering strategic thinking and social interaction. Whether in the online or offline mode, they offer a diverse spectrum of experiences, engaging players in skillful pursuits and immersive narratives. Today, games are also a medium of business, drawing players from all corners of the world and impacting the economy. For this sunrise industry, the tech fraternity has also given their vote of confidence (this is part of the findings of Mukherjee and Maitra, 2023). Ludo is a popular board game for 2-4 players, originated in India and inspired by the ancient game of Pachisi. The British popularized it globally in the early 20\textsuperscript{th} century. In this paper we will attempt a detailed empirical analysis of a couple of variants on the Ludo Game that give rise to diverse strategic implications in the game play. \footnote{Tomasev et.al. (2020) highlight the strategic changes of new variants of a pre-existing game using the example of chess.} The subsections \ref{ludo_desc_2p} and \ref{ludo_desc_4p} in the next section describes these two variants. 

\subsection{Game Description: Generic Ludo}
\label{ludo1}
In the game of ludo, each player rolls a die and moves their four tokens around the board, trying to get them to the destination squares. The tokens are usually of four different colours: \textit{red}, \textit{green}, \textit{yellow} and \textit{blue}. One token is considered to be active only after a 6 is rolled, and the player can start moving the token after that. One can capture others' tokens, which will lead the captured token to be considered inactive again. All four tokens are initially inactive. The first person to get all their tokens to the destination square, which is 56 squares away from the start, wins.

Once activated, tokens start from their respective coloured squares adjacent to the 6x6 corner squares, move around the board and aim to reach the finish, which is the purple square adjacent to their coloured row of squares in the 8\textsuperscript{th} row. Thus, (7, 2) is the starting square for red tokens and (9, 14) is the starting square for blue tokens. Again (8, 7) is the destination square for the red tokens and (8, 9) is the destination square for the blue tokens. 

The game starts off with alternate dice rolls by the four players. When a 6 appears in roll of any player, one of their tokens in considered to become active and moves to the beginning square. In the next turn, if a 6 appears, either the second token can be  activated or the first token can be moved 6 squares; otherwise, the first token moves a number of squares equal to the roll, in it's designated path. In any of the consequent turns, if 6 appears, an inactive token may be activated as long as there are such tokens. If a red token reaches the same position as blue token (except their respective starting squares) after the roll, the blue token is 'captured' and considered inactive, and vice-versa. No token can be captured on the starting squares of any player or one of it's safe squares (the row of squares with the same colour on the 8\textsuperscript{th} row.)

As long as there are zero or one active tokens, there is only one choice of play available to any player. However, choices and complexities arise when there are multiple active tokens. \\

{\bf Extra move rule:}

In ludo, a player gets an extra ‘dice roll’ every time they either (a)
roll a six, (b) cut another player’s token or (c) get a token home. 
\begin{itemize}
    \item  In both the formats, in the event a player simultaneously rolls a six and
either (1) cuts another player’s token or (2) gets a token home, they will
only get one extra turn.
    \item  If a player rolls 3 sixes consecutively, the third six rolled is treated as
null and void. Hence, if a player rolls three sixes in a single turn, the third
six will not count and only two sixes will count for that turn. Upon rolling
third six, the player will simply pass the turn to the next player.
\end{itemize}

\subsection{First New Variant of the Game: 2-player, 3-dice version}
\label{ludo_desc_2p}
Now, let us consider a new board game, played on a board similar to a ludo board, but involving two players with three independent die and a fixed number of turns.

In the first turn of the game, the first moving player (for example, red) rolls the three die. He can choose any of the three rolls and move any of his tokens by a number of squares equal to that roll. It is to be noted that all tokens are considered active at the beginning of the game, with no requirement for 6 to start. Next, the second moving player (say, blue), chooses one of the die and moves any of their tokens. The third roll, by default, is given to red. 

In the next turn, blue rolls the three die and chooses any of those to move his tokens. Red then chooses one of the other two and the last one is given to blue. This continues till the total number of turns, considered to be either 16, 20 or 24, is reached (we consider results for all of these options). Thus, the total number of moves made by either player (excluding extra moves), is 24, 30 or 36 respectively.

The players still get an extra turn on rolling a six, capturing a token, or getting a token home. However, that extra turn is not considered to be a separate turn. When two or more tokens of same colour are on a square, they form a safe zone and cannot be captured by opponent tokens reaching that square. 

Rules of promotion and capturing are similar as in generic ludo,  but the difference lies in the win criteria, which is based on a point system unlike generic ludo. 
A player is awarded 1 point for each square traveled by a token. Further, an extra 56 points are awarded for getting the token to "home". If a token is captured, it's points accrued are removed. At the end of the game, the player with more number of points wins the game.

\subsection{Second New Variant of the Game: 4-player, 5-dice version}
\label{ludo_desc_4p}
Now, let us consider a new board game, played on a board similar to a ludo board, but with five independent die and a fixed number of turns.

In the first turn of the game, the first moving player (for example, red) rolls the five die. He can choose any of the five rolls and move any of his tokens by a number of squares equal to that roll. It is to be noted that all tokens are considered active at the beginning of the game, with no requirement for 6 to start. Next, the second moving player (say, blue), chooses one of the die and moves any of their tokens. Next, the third moving player (say, yellow), chooses one of the three available die and moves any of their token, followed up by the fourth moving player(say, green) who performs a similar action with the two available die. For his second move, red moves any token with the last remaining die. 

The second turn starts with the second moving player, blue, rolling the five die again and choosing one of them to move a token. Next, yellow chooses one of the four available die to move one of their tokens, followed by green, red and blue who perform similar actions in that order.

The third turn is played in a similar manner, with yellow, green, red, blue and yellow making their moves with one of the available die in that order, with yellow making their first move of the turn with five available die.

The fourth turn is played in a similar manner, with green, red, blue,yellow and green making their moves with one of the available die in that order, with green making their first move of the turn with five available die.

This continues till the total number of turns, considered to be either 8, 12 or 16 is reached. Thus, the total number of moves made by each player(excluding extra moves) is 10,15 or 20 respectively. We consider results for all these variations in this study.

The players still get an extra turn on rolling a six, capturing a token, or getting a token home. However, that extra turn is not considered to be a separate turn. When two or more tokens of same colour are on a square, they form a safe zone and cannot be captured by opponent tokens reaching that square. 

The point system in this version is similar to the 2-player, 3-dice variant.
A player is awarded 1 point for each square traveled by a token. Further, an extra 56 points are awarded for getting the token to "home". If a token is captured, it's points accrued are removed. At the end of the game, the player with more number of points wins the game.

In the rest of the paper, the analysis proceeds as follows. Section \ref{theory} outline the (Nash) equilibrium calculations that we use in the paper. The section \ref{stat} describes the specification of players' Rule-based Strategies for empirical evaluation. These evaluations are listed sequentially in section \ref{empirical} over several subsections. The resulting Nash equilibria for different variants are discussed in section \ref{nash} along with a comparative discussion.

\section{Nash Equilibrium on Expected Path}
\label{nash}
\label{theory}
Before embarking on a detailed empirical analysis, let us discuss what the game structure tells us on paper in terms of expected outcome(s). For the sake of simplicity, we will assume the expected point 3.5 $(= \frac{1+2+3+4+5+6}{6})$ in each throw and a Nash equilibrium\footnote{Nash equilibrium is a very intuitive and the most popular notion of equilibrium in strategic games. For detailed discussion, refer to Dutta (1999). It suffices to say that a pair of Nash equilibrium strategies for a two player game is one where no player can benefit by deviating unilaterally.} consideration for strategy selection by the players. Similar interpretation can be done for four player games as well. As playing according to a strategy is in itself an act of skill, a favourable outcome with strategic playing supports the skill base of any game, that involves uncertainty (nature's move) over chance. It is pertinent to mention that we are considering only a limited number of simple and intuitive strategies to understand the role of skill (here strategic behaviour) in winning this game. There could be many other, more complex strategies possible. Here an initiating discussion is presented, and more comprehensive statistical results with different strategies will be explained in the following sections. 

\subsection{2-player, 3-dice version}

First note the following for the 24 move Ludo game: 

(a)	Expected moves \& points: 
\begin{itemize}
\item 24 moves 
\item $ + 4$ moves as one 6 is expected in six moves 
\item $ + \frac{2}{3}$ moves as a double six is expected in thirty-six moves, 
\item each getting 3.5 on average for $24+4=28$ moves, except for the double 6, where it is 2.5 (as the third six will be spoiled, in this case the average is $\frac{1+2+3+4+5}{6} = 2.5$). 
\item That is, expected number of total points $= 28 \times 3.5 + \frac{2}{3} \times 2.5  = 99.67$ 
\end{itemize}
(b) Pieces have symmetric chances to capture each other (hence $\frac{1}{2}$ each). Thus, the average progress will be half the remaining path if there are opposing pieces moving in the same region.

The strategies are defined in terms of pairs of players: \{R1, R2\}, \{R3, R4\} and \{B1, B2\}, \{B3,  B4\} as it is easy to show that moving in pairs with a little staggering (small gap) is preferable in obtaining protection to moving alone. It can be noted that the first 12 moves (50\% of the game) will roughly consist of reaching the opposition's starting point. After this, the following strategies are considered for the remaining 12 moves: 
\begin{enumerate}
\item Promotion priority (PP): The leading two pieces {R1, R2} keep moving forward and tries to get one of them (w.l.g. R1) to finish (moving 56 steps + getting 56 bonus points).
\item Safe progress priority (S): The other two pieces {R3, R4} start moving and tries to reach the same point as {R1, R2}, so no attempt at finishing.
\item Mix of the two strategies above (M): Move R1 and R3 with the objective of reaching finish for R1 and progress for R3.
\end{enumerate}
We consider expected outcomes when each possible pair of strategies play against each other. This is calculated assuming uniform probability of capture. The payoffs are calculated as below.

\begin{itemize}
    \item Strategy (PP) vs. (PP): $(99.67 + 56 + 3.5, 99.67 + 56 + 3.5).$ This is total expected number of points from movements plus the 56 bonus points for promoting one coin. While this may have a second order effect of getting another six in the extra move and then getting some more points. As these come with very small probability and it will not change our results qualitatively, we ignore this for the present discussion.
    \item Strategy (S) vs. (S): $(99.67, 99.67)$ (in this case, there is no bonus as promotion does not occur).
    \item Strategy (PP) vs. (S)\\
Payoff for (PP): $\frac{1}{4} (99.67 + 56 + 7) + \frac{1}{2} (37 + 12.5 + 3.5) + \frac{1}{4} (12 + 12) = 73.$ \\
The above calculation is based on four possibilities where the first player captures (or does not capture) times the second player captures (or does not capture). Assuming equi-probability, these occur with $\frac{1}{4}$, $(\frac{1}{4} + \frac{1}{4} =) \frac{1}{2}$ and $\frac{1}{4}$ probability respectively. When both sides do not capture, then there is expected movement ($99.67$) plus promotion bonus plus the gain from extra moves won for the capture. In the adverse case, with both sides capturing, the token is able to achieve a net movement of half the length of his home stretch only. In the mixed case, the progress is full home stretch plus half enemy territory for one token, half of home stretch for the second one plus one bonus move for capture.
\\
Payoff for (S): $50 + \frac{1}{4} (12 + 12) + \frac{1}{2} (37 + 3.5) + \frac{1}{4} (50 + 7) = 90.5$ \\ The calculation use the same logic as above.
    \item Strategy (M) vs. (M): $\frac{1}{4} (56 + 56 + 19) + \frac{1}{4} (12.5 +12.5) + \frac{1}{4} (56 + 56 + 9.5) + \frac{1}{4} (12.5 + 25) = 103$ \\ Now, as the promotional possibility is pursued, in the best - no capture - case we get 56 plus bonus 56 and the remaining expected progress. This happens with probability $\frac{1}{4}$. The worst case of both capture - again with probability $\frac{1}{4}$ - results in scores of 12.5 expected for each of the two tokens used. The mixed case of one promotion and one capture gets 56 plus bonus 56 and half of the remaining expected progress. The other mixed case with only the advance token getting captured is similarly computed.
    \item Strategy (M) vs. (S) \\
Payoff for (M): $25 + \frac{3}{4} (12.5 + 25) + \frac{1}{4} (30 + 56 + 19 + 3.5) = 80.25$ \\ Again, the probability of capture is attributed through the possible combinations. Now (M) suffers a higher risk of being captured (in 3 out of 4 possible configurations, hence $\frac{3}{4}$) compared to the (S) player.  \\
Payoff for (S): $50 + \frac{3}{4} (50 + 3.5) + \frac{1}{4} (12) = 93$
\item Strategy (M) vs. (PP) \\
Payoff for (M): $25 + 30 + 56 + \frac{3}{4} (10 + 3.5) + \frac{1}{4} (19) = 126$ \\
Payoff for (PP): $\frac{3}{4} (99.67 + 56) + \frac{1}{4} (12) = 120$ \\ Again, similar asymmetry of capture probability is present between (M) and (PP). We omit these details as the logic is similar to the earlier cases.
\end{itemize}

We summarise these and the corresponding outcome (payoff = 1(W) or 0(L)) in the tables below: \\

\begin{table}[H]
\centering
\begin{tabular}{|c|c|c|c|}
\hline
points &	(PP)	& (S)	& (M) \\
\hline
(PP)&	(159.17, 159.17)&	(73, 90.5)&	(120, 126) \\
(S)&	(90.5, 73)&	(99.67, 99.67)	& (93, 80.25) \\
(M)&	(126, 120)	& (80.25, 93)&	(103, 103) \\
\hline
\end{tabular} 
\caption{Points on Expected path for 16-turn game}
\end{table}

\begin{table}[H]
\centering
\begin{tabular}{|c|c|c|c|}
\hline
Payoff (W/L) &	(PP)	& (S)	& (M) \\
\hline
(PP)&	(1/2, 1/2) &	(0, 1)&	(0, 1) \\
(S)&	(1, 0)&	{\bf (1/2, 1/2)}	& (1, 0) \\
(M)&	(1, 0)	& (0, 1)&	(1/2, 1/2) \\
\hline
\end{tabular} 
\caption{Payoff on Expected path for 16-turn game}
\end{table}

\subsection{4-player, 5-dice version}
First note the following for the 16 turn, 20 move Ludo game: 

(a)	Expected moves \& points: 
\begin{itemize}
\item 20 moves 
\item $ + \frac{10}{3}$ moves as one 6 is expected in six moves 
\item $ + \frac{5}{9}$ moves as a double six is expected in thirty-six moves, 
\item each getting 3.5 on average for $20+\frac{10}{3}=\frac{70}{3}$ moves, except for the double 6, where it is 2.5 (as the third six will be spoiled, in this case the average is $\frac{1+2+3+4+5}{6} = 2.5$). 
\item That is, expected number of total points $= \frac{70}{3} \times 3.5 + \frac{5}{9} \times 2.5  = 83.06$ 
\end{itemize}
(b) Pieces have symmetric chances to capture each other (hence $\frac{1}{2}$ each). Thus, the average progress will be half, one-fourth or one-eighth the remaining path if there are one,two or three sets of opposing pieces moving in the same region.

The strategies are defined in terms of pairs of players: \{R1, R2\}, \{R3, R4\} and \{B1, B2\}, \{B3,  B4\} as it is easy to show that moving in pairs with a little staggering (small gap) is preferable in obtaining protection to moving alone. It can be noted that the first 5 moves (25\% of the game) will roughly consist of reaching the end of it's home stretch. After this, the following strategies are considered for the remaining 15 moves: 
\begin{enumerate}
\item Promotion priority (PP): The leading two pieces {R1, R2} keep moving forward and tries to get one of them (w.l.g. R1) to finish (moving 56 steps + getting 56 bonus points).
\item Safe progress priority (S): The other two pieces {R3, R4} start moving and tries to reach the same point as {R1, R2}, so no attempt at finishing. Once all pieces have reached the same point, they start moving in pairs again, aiming to reach the next safe square this time.
\item Mix of the two strategies above (M): Move R1 and R3 with the objective of reaching finish for R1 and progress for R3.
\end{enumerate}
We consider expected outcomes when each possible quartet of strategies play against each other. For simplicity, we start off with combinations of S and PP only, noting there are $2^4 = 16$ combinations here, instead of $3^4 = 81$ combinations that would arise from including M.(Since the Nash equilibrium in 2 player case does not include M, and it is a mixture of the two extreme strategies PP and S).This is calculated assuming uniform probability of capture. The payoffs are calculated as below.

\begin{itemize}
    \item Strategy combination (PP, PP, PP, PP): $(83.06 + 56 + 3.5, 83.06 + 56 + 3.5,83.06 + 56 + 3.5, 83.06 + 56 + 3.5).$ = $(132.56, 132.56,132.56,132.56)$ This is total expected number of points from movements plus the 56 bonus points for promoting one coin. While this may have a second order effect of getting another six in the extra move and then getting some more points. As these come with very small probability and it will not change our results qualitatively, we ignore this for the present discussion.
    \item Strategy combination (S, S, S, S): $(83.06, 83.06,83.06,83.06)$ (in this case, there is no bonus as promotion does not occur).
    \item We note that  
    Strategies (PP, S, S, S), (S, PP, S, S), (S, S, PP, S) and (S, S, S, PP) are similar with respect to the relative position of players( some differences might arise due to the absolute position of the PP player), so they can be analyzed together, accounting for the PP player's position separately.
Payoff for (PP): $\frac{1}{16} (83.06 + 56 + 3.5) + \frac{1}{16} (83.06 + 56 + 7) + \frac{7}{16} (32.5 + 13 + 3.5) + \frac{7}{16} (26) + x$= $50.85 + x$ (where x is a variable depending on the position of PP player).
The above calculation is based on four possibilities where the first player captures (or does not capture) times the any of the other players captures (or does not capture). Assuming equi-probability, these occur with $\frac{1}{16}$, $\frac{1}{16}$, $\frac{7}{16}$ and $\frac{7}{16}$ probability respectively. When both sides do not capture, then there is expected movement ($83.06$) plus promotion bonus plus the gain from extra moves won for the promotion. In the adverse case, with both sides capturing, the token is able to achieve a net movement of the length of his home stretch only. In the mixed case, the progress is around 2.5 times the home stretch for one token, half of home stretch for the second one plus one bonus move for capture. Indeed, there are differences in probabilities of captures by other players( for e.g, red travels more squares overlapped by blue than green), but this is approximately balanced by the fact red gains less points on an average when captured by blue than green.

Now, the PP player takes an expected number of $\frac{13*2}{3.5} \approx 7$ turns before it moves into the next player's territory. Further, the 8th turn starts with the 1st, 2nd, 3rd and 4th  players getting 3,2,1 and 5 rolls respectively. So, the expectation of PP player can be changed by a factor of $\frac{2}{5}*(3.5)$, $\frac{1}{5}*(3.5)$,$0$ and $\frac{4}{5} *(3.5)$ respectively, which is the product of an approximate increase in capture probability(since all usual capture calculation was done considering one dice, and here both PP and S player more than one dice available. Captures are also possible in later turns, but that probability is even lower, and is hence ignored) and points gained in by extra move due to capture. This factor is x.

While considering the other two S players,it is easy to see that the probability of PP player meeting them in a 20-turn game is negligible, so x = 0. 

Payoff for (S): $52 +  \frac{1}{4}{6*3.5 + 3.5} + \frac{1}{4}{3*3.5} + \frac{1}{2}{3*3.5 + 3.5} + y$(where y is a variable depending on the position of PP player).

Now,it is assumed that S players don't capture each other, so capture possibilities only occur if S player moves immediately after the PP player.

For that S player, y = -x , where the added factor is expected number of squares moved by S player depending on whether S and PP player capture(or do not capture).

The other two S players interact with PP player with negligible probability, so their expected points is simply 83.06.

    \item  Strategies (PP, PP, S, S), (S, PP, PP, S), (S, S, PP, PP) and (PP, S, S, PP) are similar with respect to the relative position of players( some differences might arise due to the absolute position of the PP players), so they can be analyzed together, accounting for the PP player's position separately.

    We first note that the two PP players interact with negligible probability, so it is assumed that they don't capture each other. S players show similar behaviour.
    Also using previous logic, the two S players can be assumed to not interact with the first PP player, and the second PP player can be assumed to interact with only the S player moving immediately after it. 
    
    The expected points for the PP and S players that interact are exactly similar as the calculation in the one PP, 3 S player case. 
    
\item  Strategies (PP, S, PP, S) and (S, PP, S, PP) are similar with respect to the relative position of players( some differences might arise due to the absolute position of the PP players), so they can be analyzed together, accounting for the PP player's position separately. 

Since we have already seen PP and S players not adjacent to each other interact with negligible probability, these quartets can be considered to be two PP-S pairs. The payoffs of the PP and S players in these pairs can be calculated similar to the adjacent PP-s pair in one PP, 3 S case. 

\item   Strategies (PP, PP, PP, S), (S, PP, PP, PP), (PP, S, PP, PP) and (PP, PP, S, PP) are similar with respect to the relative position of players( some differences might arise due to the absolute position of the S player), so they can be analyzed together, accounting for the S player's position separately.

Since PP players don't interact with each other or with S players(if not moving immediately after PP player), two of the PP players(not moving immediately before the S player), have same expectation. 
The expected points for the remaining PP-S pair is exactly similar to the earlier cases. 

\end{itemize}


The strategies considered in the empirical analysis will be refinements of these, subject to various possibility of capture. Thus, when we analyse the game-play strategies in simulation, we will keep track of the distribution of points achieved.

While we understand that the complexities of various game play paths are too high to be fully addressed in this study, we do try to visualise these in terms of some rather simplified statistics like win percentage, average points scored and its variability. While winning is the ultimate goal, the points gained even in a loss has important signals about the nature of game play. A high risk strategy may lose against a more safe strategy often but still score points at a higher rate (losing closely and winning hands down, but less often). This will then show up in the variability measure (here SD). The skill of a player then would show up in being able to choose the right (combination of) strategy given the version of the game and may as well depend on the state of play. We explore these issues in a limited way below.

\section{Specification of Players' Intuitive Rule-based Strategies for Empirical Evaluation}
\label{stat}
For the empirical evaluation, we consider three intuitive rule-based strategies and compare the performances of players following all possible combinations of these strategies(using a frequentist approach, as outlined by Mukherjee,Maitra and Das(2023)). In this section, we describe these strategies in detail. \\ 
We use three alternatives where one is similar to (PP) (called "aggressive"), a second similar to (S) (called ``responsible pair") and a relatively unsophisticated algorithm (called "naive"). We detail below.

\subsection{The ``Naive" Player (N)}

The naive player follows a very simple algorithm for movement of tokens.  As long as the first token is active, the token is moved by a number of squares equal to the first available die roll in it's designated path; unless the move is illegal because it crosses it's destination square with that move. \\ If the first token is inactive or it's movement is illegal, the next active token is moved in a similar way. \\
If the first token is captured at any point, it is still given priority in movement compared to the other tokens. That is, the capture of a token does not affect it's movement algorithm when it's activated in the future.
\\ All tokens are considered inactive when they reach their destination, and the player wins if all his tokens reach the destination before the opponent's.

\subsection{The ``Aggressive" Player (A)}

The aggressive player follows an algorithm similar to the unskilled player's , but with a few more choices, which usually work in the aggressive player's favour.\\
The main difference lies in the choice of movement of an active token for the aggressive player. If there is the possibility of capturing an opponent token with a move, the aggressive player prioritizes that capture (for example, say an opponent token is 4 squares away from the third player token, and the roll is 4. Then the aggressive player will prioritize moving the third token over the other actives and capture the opponent token.) \\ Further, if it is observed that one of the player tokens can reach the 'safe' squares (the row of same coloured squares in the 8\textsuperscript{th} row, or the starting squares of any opponent, where the opponent tokens can't capture it) after the turn, it is given priority and moved off to the safe square. However, if a token can be promoted, i.e, moved to it's destination square, it is given the highest priority, since it will award at least 57 points to the player. \\
If none of the above movements are possible, the aggressive player moves the first active token, choosing the highest roll available.  
Similar to the "naive" player, the capture of a token doesn't affect it's movement algorithm. \\
Thus, the action performed by the aggressive player in any specific turn (if possible), in decreasing order of priority is-
\begin{itemize}
\item Promoting a token
    \item Capturing an opponent token
    \item Moving a token to one of the safe squares close to promotion
    \item Moving the first active token with the highest roll available
\end{itemize}

\subsection{The ``Responsible Pair" Player (RP)}

The player following the "Responsible pair" strategy has a considerably more complicated algorithm, initially aiming to get all of his tokens to the diagonally opposite opponent's starting square. 
The first two tokens start moving in alternate turns, and once either has reached 27, one of the two remaining tokens start moving in their place. After all tokens have reached 27, the first two tokens start moving in alternate turns.

After crossing the 27$^{th}$ square, all tokens prioritize captures in their movements. However, before a token has crossed the 27$^{th}$ square, it cannot cross that square to make a capture. 

When an opponent token is within a distance of six (the maximum dice roll) of one of the last two tokens of the player, that token is given priority in movement, in an attempt to capture the opponent token.

When one of the opponent's tokens have entered the safe squares close to the destination and are ready to promote, the "responsible pair" player prioritizes the movement of their highest value token, aiming to promote it. All other tokens are only used for captures and moving to safe squares if possible. If that token gets captured, the next highest value token is given priority, 

At any move, if no skilled movement is possible, the responsible pair player chooses the highest roll available.

Thus, the actions that can be performed by the "responsible  pair" player, in decreasing order of priority, are-
\begin{itemize}
    \item Promoting a token
    \item Capturing an opponent token
    \item Moving a token to a safe square
    \item Moving the highest point token when the opponent's token is close to promotion
    \item Chasing an opponent token with either of it's last two tokens
    \item Moving the tokens in alternate turns till none of them have reached 27, provided none of them reach 27 with the highest feasible roll
    \item Moving the tokens in alternate turns after all of them have reached 27 with the highest feasible roll
\end{itemize}

As the four player variant has a much higher intensity of interaction among pieces belonging to different players, we consider shorter games in our empirical evaluation that follows. Three variants of the game depending on the number of moves allowed are considered, consisting of 8, 12 and 16 total turns respectively. We simulated the games 1000 times for each strategy combination and the results have been presented in the following section.

\section{Empirical Results}
\label{empirical}

This sections lists all the empirical results for the two variants along with different lengths of the game considered.

\subsection{2-player, 3-dice version}

\begin{center}
    {\bf 16-turn, 24-move game}
\end{center}

\begin{table}[H]
\small
\centering
\begin{tabular}{|c|c|c|c|c|c|c|c|}
\hline
\multirow{2}{2 cm}{\textbf{Player 1 Strategy}} & \multirow{2}{2 cm}{\textbf{Player 2 Strategy}} & \multirow{2}{3 cm}{\textbf{Player 1 Win Percentage}} 
& \multicolumn{2}{c|}{\textbf{Player 1 points}} & \multicolumn{2}{c|}{\textbf{Player 2 points}} \\
\cline{4-7}
& & & \textbf{Mean} & \textbf{SD} & \textbf{Mean} & \textbf{SD} \\
\hline
Naive & Naive & 49.83 & 159.7 &  41.01 & 157.2& 42.39 \\
\hline
Naive & Aggressive & 0.22 & 75 &53.16& 209.6 & 48.65 \\
\hline
Naive & Responsible Pair & 13.6 & 82.2 &54.1& 135.1 & 39.51 \\
\hline
Aggressive & Naive & 97.07 & 211.5 &48.67& 74.7& 52.83 \\
\hline
Aggressive & Aggressive & 48.12 & 120.6 &59.06& 121.9 & 58.57 \\
\hline
Aggressive & Responsible Pair & 44.42 & 114.5 &71.78& 109.3 & 44.37 \\
\hline
Responsible Pair & Naive & 83.6 & 133.6 &38.77& 86.6 & 54.09 \\
\hline
Responsible Pair& Aggressive &49.88&105.3&43.59&117.7&69.59\\
\hline
Responsible Pair& Responsible Pair&49.68&99.1&16.02&99.3&16.31\\
\hline
\end{tabular}
\caption{Summary Statistics for various strategies in 16-turn game}
\label{ludo_2p_16turns}
\end{table}

\begin{center}
    {\bf 20-turn, 30-move game}
\end{center}

\begin{table}[H]
\small
\centering
\begin{tabular}{|c|c|c|c|c|c|c|c|}
\hline
\multirow{2}{2 cm}{\textbf{Player 1 Strategy}} & \multirow{2}{2 cm}{\textbf{Player 2 Strategy}} & \multirow{2}{3 cm}{\textbf{Player 1 Win Percentage}} 
& \multicolumn{2}{c|}{\textbf{Player 1 points}} & \multicolumn{2}{c|}{\textbf{Player 2 points}} \\
\cline{4-7}
& & & \textbf{Mean} & \textbf{SD} & \textbf{Mean} & \textbf{SD} \\
\hline
Naive & Naive & 49.68 & 210.2 &50.31& 209.1& 51.66 \\
\hline
Naive & Aggressive & 1.3 & 99.4 &61.64& 269.9 & 50.9 \\
\hline
Naive & Responsible Pair & 13.01 & 125.1 &61.73& 194.4 & 53.91 \\
\hline
Aggressive & Naive & 98.19 & 268.4 &52.32& 100.1& 60.98 \\
\hline
Aggressive & Aggressive & 48.69 & 159.6 &69.49& 160.8 & 69.9 \\
\hline
Aggressive & Responsible Pair & 50.06 & 162.4 &78.72& 148 & 52.93 \\
\hline
Responsible Pair & Naive & 84.97 & 191.2 &52.7& 130.4 & 61.53 \\
\hline
Responsible Pair& Aggressive &47.09&143.4&50.61&164&76.29\\
\hline
Responsible Pair& Responsible Pair&48.49&128.3&28.7&129.4&29.27\\
\hline
\end{tabular}
\caption{Summary Statistics for various strategies in 20-turn game}
\label{ludo_2p_20turns}
\end{table}

\begin{center}
    {\bf 24-turn, 36-move game}
\end{center}

\begin{table}[H]
\small
\centering
\begin{tabular}{|c|c|c|c|c|c|c|c|}
\hline
\multirow{2}{2 cm}{\textbf{Player 1 Strategy}} & \multirow{2}{2 cm}{\textbf{Player 2 Strategy}} & \multirow{2}{3 cm}{\textbf{Player 1 Win Percentage}} 
& \multicolumn{2}{c|}{\textbf{Player 1 points}} & \multicolumn{2}{c|}{\textbf{Player 2 points}} \\
\cline{4-7}
& & & \textbf{Mean} & \textbf{SD} & \textbf{Mean} & \textbf{SD} \\
\hline
Naive & Naive & 49.49 & 259.3 &53.49&  257.8 &54.2 \\
\hline
Naive & Aggressive & 0.95 & 128.5&68.96 &331.3 & 55.15 \\
\hline
Naive & Responsible Pair & 13.25 & 168.8&65.66 &262.3& 65.99 \\
\hline
Aggressive & Naive & 98.93 & 330.2&56.23 &126& 68.43 \\
\hline
Aggressive & Aggressive & 49.31 & 196.6&77.5 &196.9& 77.49 \\
\hline
Aggressive & Responsible Pair & 58.99 & 213& 84.53 &181.8 & 55.93 \\
\hline
Responsible Pair & Naive & 84.1 & 260.1&67.04 &173.8& 65.27 \\
\hline
Responsible Pair& Aggressive & 35.89 &177.3&53.83&216.7&82.24\\
\hline
Responsible Pair&Responsible Pair&48.77&189.9&57.99&191.4&57.15\\
\hline
\end{tabular}
\caption{Summary Statistics for various strategies in 24-turn game}
\label{ludo_2p_24turns}
\end{table}

\subsection{4-player, 5-dice version}

\begin{center}
 {\bf 16-turn, 20-move game}  
\end{center}

\begin{longtable}[H] {|p{0.4cm}|p{0.55cm}|p{0.55cm}|p{0.55cm}|p{0.55cm}|p{0.75cm}|p{0.75cm}|p{0.75cm}|p{0.75cm}|p{0.9cm}|p{0.8cm}|p{0.9cm}|p{0.8cm}|p{0.9cm}|p{0.8cm}|p{0.9cm}|p{0.8cm}|}
\hline
Sl. No. & \multicolumn{4}{c|}{Strategies} & \multicolumn{4}{c|}{Win Percentages} & \multicolumn{2}{c|}{P1 Points} & \multicolumn{2}{c|}{P2 Points} & \multicolumn{2}{c|}{P3 Points} & \multicolumn{2}{c|}{P4 Points} \\ 
\hline
 & P1 & P2 & P3 & P4 & P1 & P2 & P3 & P4 & Mean & Sd & Mean & Sd & Mean & Sd & Mean & Sd \\
\hline
\endhead
\hline
\endfoot
1&N&N&N&N&26.22&24.58&24.58&24.62&108.2&49.4&105.66&51.59&106.36&49.97&108.95&48.89\\
2&A&N&N&N&89.25&2.4&4.3&4.05&183.89&59.72&37.05&36.67&53.38&49.17&65.61&52.74\\
3&RP&N&N&N&55.8&11.25&15.15&17.8&114.11&40.47&66.92&49.3&55.35&46.64&57.72&49.26\\
4&N&A&N&N&4.4&89.45&2.65&3.5&66.08&53.16&189.58&57.99&41.16&41.4&51.81&48.18\\
5&A&A&N&N&63.2&33.3&1.2&2.3&137.19&62.2&96.39&64.72&22.15&24.4&29.43&33.76\\
6&RP&A&N&N&33.7&53.4&5.9&7&90.43&52.58&115.94&75.92&28.54&30.08&32.9&36.69\\
7&N&RP&N&N&19&50.75&15.8&14.45&61.66&51.26&112.22&38.35&73.02&51.59&55.55&46.61\\
8&A&RP&N&N&73.6&17.9&2.8&5.7&132.67&65.96&73.43&25.3&28.81&32.02&34.8&39.46\\
9&RP&RP&N&N&23.45&55.7&8.35&12.5&68.92&24.19&84.55&27.43&35.7&37.36&36.16&38.49\\
10&N&N&N&A&1.45&2.4&3.1&93.05&39.11&39.89&52.47&49.85&64.66&52.07&192.54&57.09\\
11&A&N&N&A&27.6&1.15&3.15&68.1&93.15&63.74&19.97&23.14&29.54&37.45&142.4&60.65\\
12&RP&N&N&A&19.05&2.5&3.2&75.25&73.97&25.84&27.74&34.03&28.01&33.93&137.58&66.84\\
13&N&A&N&A&3.8&44.45&3.5&48.25&24.36&31.25&112.38&65.32&21.92&27.01&113.13&63.49\\
14&A&A&N&A&33.55&25.05&2.7&38.7&65.59&56.57&56.08&54.75&13.22&15.89&76.71&61.34\\
15&RP&A&N&A&25.25&27.6&2.8&44.35&54.46&39.54&64.25&58.93&15.95&20.01&79.87&61.94\\
16&N&RP&N&A&2.9&26.35&4.55&66.2&24.04&28.32&80.86&27.05&32.34&36.5&122.71&68.6\\
17&A&RP&N&A&28.35&28.1&2&41.55&64.54&57.47&61.84&23.3&16.37&22.27&83.27&61.71\\
18&RP&RP&N&A&10.3&36.05&1.7&51.95&50.76&20.98&68.37&22.31&17.32&22.67&85.76&61.93\\
19&N&N&N&RP&8.6&13.15&18.65&59.6&63.41&49.22&49.94&46.63&60.27&49.94&117.21&41.91\\
20&A&N&N&RP&50.55&4.2&6.6&38.65&113.6&72.9&26.41&31.19&31.39&37.59&95.75&56.36\\
21&RP&N&N&RP&55.65&6.7&10.6&27.05&85.67&26.9&33.29&36.14&32.2&35.57&72.13&24.56\\
22&N&A&N&RP&3.6&63.2&3.3&29.9&30.9&37.2&118.38&69.46&25.25&29.36&83.46&29.02\\
23&A&A&N&RP&36.8&26.75&1.95&34.5&79.66&63.97&60.5&55.27&14.75&17.48&66.92&47.66\\
24&RP&A&N&RP&23.25&35.4&3&38.35&58.62&43.4&74.54&64.26&17.52&21.17&59.96&28.32\\
25&N&RP&N&RP&6.4&40.7&8.35&44.55&35.89&37.79&90.99&29.12&38.65&39.21&92.64&28.74\\
26&A&RP&N&RP&36.45&39.3&3.2&21.05&79.28&65.17&69.01&21.65&19.1&24.35&67.93&44.25\\
27&RP&RP&N&RP&17.25&56.75&5.1&20.9&55.01&21.08&74.15&22.6&21.45&26.51&58.15&21.89\\
28&N&N&A&N&3.1&4.1&89.55&3.25&51.09&49.36&66.5&53.26&184.01&60.81&44.34&41.05\\
29&A&N&A&N&45.55&2.7&49.35&2.4&116&68.4&22.21&26.29&115.98&67.77&23.94&27.08\\
30&RP&N&A&N&28.2&4.9&62.9&4&82.65&27.98&34.53&40.89&117.03&66.43&28.84&30.18\\
31&N&A&A&N&2.9&61.75&33.2&2.15&29.36&37.29&132.27&63.86&94.89&63.46&23.9&25.35\\
32&A&A&A&N&41.65&31.35&25&2&80.36&63.54&65.26&57.18&57.18&52.14&14.74&15.09\\
33&RP&A&A&N&30.85&35.1&31.75&2.3&61.64&43.25&77.07&63.61&64.33&56.22&17.31&17.59\\
34&N&RP&A&N&8.45&34.15&53.25&4.15&32.54&39.84&89.3&55.17&113.17&73.41&28.37&29.78\\
35&A&RP&A&N&46.15&22.95&28.75&2.15&85.33&63.22&55.63&39.83&67.99&62.2&17.59&18.34\\
36&RP&RP&A&N&34.98&25.93&36.03&3.05&57.01&29.19&59.69&43.91&74.21&64.11&19.03&20.6\\
37&N&N&A&A&2.6&2.15&58.75&36.5&21.6&27.72&26.65&34.37&133.15&63.06&99.08&63.03\\
38&A&N&A&A&25.93&1.45&35.13&37.48&54.67&51.03&11.59&13.76&72.34&58.5&71.33&58.63\\
39&RP&N&A&A&28.05&1.95&39.9&30.1&62.22&22.29&15.64&20.78&79.85&61.68&66.85&55.45\\
40&N&A&A&A&2.05&34.58&30.63&32.73&12.52&16.53&72.04&59.22&62.09&55.72&60.84&52.74\\
41&A&A&A&A&22.9&21.95&25.25&29.9&36.88&43.5&36.01&43.26&37.68&44.91&40.77&43.17\\
42&RP&A&A&A&29.7&21.45&23.1&25.75&39.77&29.14&42.73&49.56&42.03&45.07&43.99&46.07\\
43&N&RP&A&A&2.85&26.8&33.75&36.6&13.84&16.69&60.21&43.02&75.27&64.04&70.4&59.08\\
44&A&RP&A&A&19.53&33.93&22.15&24.38&38.13&43.71&39.76&28.05&43.31&48.06&43.6&46.01\\
45&RP&RP&A&A&26.2&25.65&21.65&26.5&40.86&21.79&44.51&31.06&46.71&52.69&47.89&49.27\\
46&N&N&A&RP&3.7&4.9&68.7&22.7&28.77&34.54&30.07&37.31&124.05&66.83&76.06&27.22\\
47&A&N&A&RP&28.2&1.7&40.55&29.55&64.35&58.47&15.02&18.64&75.53&61.82&58&38.77\\
48&RP&N&A&RP&41.8&2.4&44.9&10.9&70.34&21.59&17.83&23.01&74.11&59.48&52.76&20.37\\
49&N&A&A&RP&1.45&37.5&29.5&31.55&15.35&20.7&76.92&62.65&62.66&55.84&64.22&22.04\\
50&A&A&A&RP&21.45&22&20.85&35.7&43.69&49.51&39.63&46.18&38.31&44.82&42.49&28.66\\
51&RP&A&A&RP&28.3&23.15&20.5&28.05&44.49&31.78&47.37&54.37&40.96&45.79&42.74&21.73\\
52&N&RP&A&RP&2.55&19.7&36.65&41.1&16.89&23.47&63.28&42.92&77.03&64.2&70.42&21.23\\
53&A&RP&A&RP&21.5&26.85&21.9&29.75&43.89&50.67&42.43&27.65&46.01&49.52&45.92&29.02\\
54&RP&RP&A&RP&25.85&22.65&25.4&26.1&44.33&23&45.2&31.26&48.46&51.95&46.67&19.95\\
55&N&N&RP&N&14.7&20.25&52.3&12.75&55.19&47.01&63.98&52.4&111.7&38.73&68.1&51.46\\
56&A&N&RP&N&63.55&3.45&28.05&4.95&119.19&68.81&24.27&28.27&82.16&28.29&36.13&38.07\\
57&RP&N&RP&N&42.9&7.25&42.15&7.7&91.7&28.97&38.59&38.04&92.67&28.91&41.2&38.32\\
58&N&A&RP&N&4.7&74.1&17.85&3.35&29.4&36.07&129.68&64.17&74.37&24.65&30.9&31.89\\
59&A&A&RP&N&39.05&28.65&30&2.3&78.28&62.96&66.92&57.9&62.33&21.42&17.86&20.47\\
60&RP&A&RP&N&22.55&37.2&37.95&2.3&64.81&43.5&79.89&65.38&68.71&21.62&20.33&23.38\\
61&N&RP&RP&N&13.2&21.85&58.15&6.8&34.73&39.29&67.96&22.85&84.9&25.8&34.51&34.04\\
62&A&RP&RP&N&49.9&8.75&38.95&2.4&83.68&65.02&48.98&19.56&69.74&21.2&19.41&22.47\\
63&RP&RP&RP&N&18.6&14.7&62.6&4.1&56.87&21.5&54.22&21&75.77&21.49&22.07&24.76\\
64&N&N&RP&A&2.65&6.65&33.75&56.95&23.45&27.47&29.14&34.88&94.3&54.95&121.47&72.08\\
65&A&N&RP&A&27.85&2.1&30.75&39.3&57.93&55.25&13.27&16.03&61.37&43.65&79.83&65.27\\
66&RP&N&RP&A&39.55&2.4&19.6&38.45&68.04&22.4&16.65&22.16&63.16&43.68&79.85&63.81\\
67&N&A&RP&A&1.85&39.25&23.65&35.25&14.33&16.32&75.72&61.4&53.32&39.67&71.91&60.36\\
68&A&A&RP&A&22.15&20.08&30.48&27.28&40.63&45.49&38.88&45.54&40.15&28.63&46.91&48.81\\
69&RP&A&RP&A&27.45&23.85&24.55&24.15&45.43&31.59&47.82&51.03&42.94&30.11&47.6&49.69\\
70&N&RP&RP&A&3&31.4&25.05&40.55&16.99&20.51&54.09&27.58&57.67&42.61&77.65&63.64\\
71&A&RP&RP&A&22.3&26.2&24.6&26.9&43.55&47.81&41.07&21.06&41.74&32.17&50.97&50.54\\
72&RP&RP&RP&A&22.65&26.75&23.85&26.75&43.61&19.51&42.98&21.67&44.41&30.88&51.92&53.25\\
73&N&N&RP&RP&5.9&10.75&22.4&60.95&32.1&35.26&30.67&36.19&68.07&24.13&86.28&28.66\\
74&A&N&RP&RP&36.7&2.75&29.95&30.6&76.88&66.37&17.59&22.34&56.76&29.86&64.89&44.98\\
75&RP&N&RP&RP&60.75&3.9&17.35&18&76.2&22.24&20.43&26.62&55.4&21.55&56.61&21.21\\
76&N&A&RP&RP&2.45&48.45&9.7&39.4&17.86&24.37&80.36&63.5&49.43&21.05&70.37&21.73\\
77&A&A&RP&RP&22.45&21.85&25.2&30.5&45.84&51.91&41.48&46.41&40.98&22.06&45.23&30.97\\
78&RP&A&RP&RP&25.05&25.05&23.55&26.35&46.94&32.87&49.69&53.4&44.71&20.95&48.36&22.28\\
79&N&RP&RP&RP&4.3&17.75&14.35&63.6&20.39&26.36&56.15&21.7&54.01&22.26&76.92&22.99\\
80&A&RP&RP&RP&22.95&26.6&21.7&28.75&48.16&53.78&44.06&20.58&43.49&23.35&48.63&30.9\\
81&RP&RP&RP&RP&26.5&23.78&23.53&26.18&46.83&21.53&46.58&21.2&45.12&20.96&47.6&19.45\\

\hline
\label{ludo_4p_16turns}
\end{longtable}

\begin{figure}[H]
    \centering
    \includegraphics[width=0.6\textwidth]{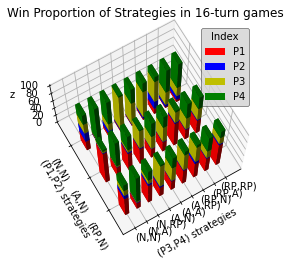} 
    \caption{Win Proportion of strategy combinations 1-27 in 16-turn game} 
    \label{fig:ludo_16turns_1-27} 
\end{figure}

\begin{figure}[H]
    \centering
    \includegraphics[width=0.6\textwidth]{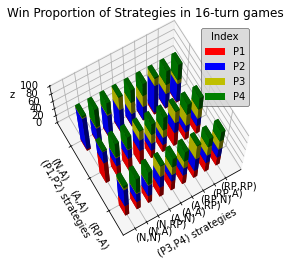} 
    \caption{Win Proportion of first strategy combinations 28-54 in 16-turn game} 
    \label{fig:ludo_16turns_28-54} 
\end{figure}

\begin{figure}[H]
    \centering
    \includegraphics[width=0.6\textwidth]{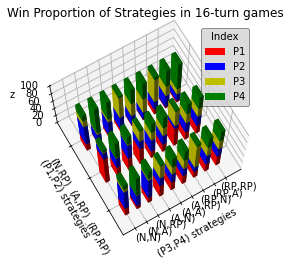} 
    \caption{Win Proportion of strategy combinations 55-81 in 16-turn game} 
    \label{fig:ludo_16turns_55-81} 
\end{figure}

\begin{center}
 {\bf 12-turn, 16-move game}  
\end{center}

\begin{longtable}[H]{|p{0.4cm}|p{0.55cm}|p{0.55cm}|p{0.55cm}|p{0.55cm}|p{0.75cm}|p{0.75cm}|p{0.75cm}|p{0.75cm}|p{0.9cm}|p{0.8cm}|p{0.9cm}|p{0.8cm}|p{0.9cm}|p{0.8cm}|p{0.9cm}|p{0.8cm}|}
\hline
Sl. No. & \multicolumn{4}{c|}{Strategies} & \multicolumn{4}{c|}{Win Percentages} & \multicolumn{2}{c|}{P1 Points} & \multicolumn{2}{c|}{P2 Points} & \multicolumn{2}{c|}{P3 Points} & \multicolumn{2}{c|}{P4 Points} \\ 
\hline
 & P1 & P2 & P3 & P4 & P1 & P2 & P3 & P4 & Mean & Sd & Mean & Sd & Mean & Sd & Mean & Sd \\
\hline
\endhead
\hline
\endfoot

1&N&N&N&N&25.05&23.6&25.8&25.55&72.43&43.32&70.32&43.23&70.32&43.03&72.25&42.49\\
2&A&N&N&N&86.75&2.8&5.25&5.2&132.72&46.09&27.52&24.83&35.53&33.85&44.79&40.27\\
3&RP&N&N&N&61.4&9.7&13.15&15.75&77.69&22.24&41.74&34.38&38.21&36.06&41.29&37.27\\
4&N&A&N&N&6.2&85.1&3.6&5.1&44.58&41.87&130.19&47.54&30.8&25.35&35.05&33.02\\
5&A&A&N&N&56.45&37.75&2.9&2.9&97.37&50.84&71.98&53.62&18.76&15.75&22.45&26.12\\
6&RP&A&N&N&33.4&53&5.3&8.3&58.61&38.25&80.68&58.79&20.47&18.36&23.64&26.79\\
7&N&RP&N&N&14.15&63.1&10.3&12.45&37.95&36.2&75.65&21.51&42.07&33.89&37.51&33.81\\
8&A&RP&N&N&64.55&30&1.9&3.55&91.58&55.54&55.62&18.24&21.24&21.73&22.65&23.45\\
9&RP&RP&N&N&28.7&57.95&4.9&8.45&50.87&20.36&62.86&17.88&24.63&24.19&25.46&26.08\\
10&N&N&A&N&3&5.15&89.85&2&32.41&33.19&44.63&39.42&134.09&43.21&28.81&22.65\\
11&A&N&A&N&49.68&3.23&43.18&3.9&80.98&57.31&17.24&18.05&77.75&54.08&18.99&15.85\\
12&RP&N&A&N&40.65&3.55&54&1.8&63.26&19.28&22.54&25.04&83.1&56.22&19.25&18.25\\
13&N&A&A&N&3.35&51.5&42.9&2.25&18.62&23.03&93.76&52.86&73.72&52.68&18.9&15.4\\
14&A&A&A&N&38.3&30.35&27.9&3.45&56.18&51.73&48.08&45.64&43.04&42.31&13.52&11.75\\
15&RP&A&A&N&31.35&37.6&28.45&2.6&42.18&26.38&55.41&49.9&45.7&44.41&14.86&13.43\\
16&N&RP&A&N&6.05&32.9&55.7&5.35&19.81&24.27&59.29&39.05&82.22&56.45&22.19&18.05\\
17&A&RP&A&N&38.93&27.75&30.08&3.23&59.07&53.49&38.82&25.58&50.63&46.96&14.49&13.61\\
18&RP&RP&A&N&36.05&25.57&34.85&3.53&42.52&21.11&40.39&25.98&55.76&51.23&16.14&14.54\\
19&N&N&RP&N&11.55&15.95&60.45&12.05&35.22&34.24&39.81&38.8&76.1&21.67&44.44&34.68\\
20&A&N&RP&N&46.95&2.85&46.9&3.3&76.93&56.08&17.87&18.21&63.7&19.26&24.04&23.17\\
21&RP&N&RP&N&43.3&4.9&47.6&4.2&68.98&18.56&24.77&26.76&69.44&19.58&26.34&24.25\\
22&N&A&RP&N&4.85&57.05&34.9&3.2&20.98&25.53&85.91&56.02&56.96&17.27&22.55&22.37\\
23&A&A&RP&N&35.75&24.95&37.95&1.35&58.46&52.84&46.47&45.02&49.02&17.48&13.79&13.04\\
24&RP&A&RP&N&22.5&30.5&45.3&1.7&45.78&26.44&52.74&49.48&53.32&17.63&16.05&15.6\\
25&N&RP&RP&N&7.7&25.55&61.7&5.05&22.12&25.67&50.3&19.36&63.25&17.52&24.71&23.7\\
26&A&RP&RP&N&38.95&13.9&44.6&2.55&58.15&52.77&38.33&16.83&53.78&17.42&16.93&16.27\\
27&RP&RP&RP&N&24.5&17.55&54.45&3.5&45.13&19.96&41.21&18.07&57.74&17.22&18.14&18.89\\
28&N&N&N&A&2.35&4.4&4.65&88.6&29.16&28.3&34.83&35.64&42.96&40.53&134.5&46.69\\
29&A&N&N&A&33.9&2.55&3.6&59.95&70.12&53.07&17.45&16.58&20.18&23.88&97.28&52.58\\
30&RP&N&N&A&31.65&2.8&4.85&60.7&57.04&18.07&20.32&21.57&23.58&26.75&90.21&55.65\\
31&N&A&N&A&3.95&43.25&4.7&48.1&16.84&19.87&79.84&55.09&18.74&19.88&79.69&53.17\\
32&A&A&N&A&29.95&24.35&4.15&41.55&49.03&47.29&42.92&44.18&11.71&12.07&59.15&49.88\\
33&RP&A&N&A&24.85&30.15&2.25&42.75&37.92&24.8&47.85&48.29&12.75&12.69&56&50.24\\
34&N&RP&N&A&1.55&42.55&3.85&52.05&17.21&17.92&62.22&19.83&23.39&24.47&81.18&55.2\\
35&A&RP&N&A&24&37.55&1.75&36.7&45.97&44.61&49.05&18.72&13.41&15.17&60.05&53.47\\
36&RP&RP&N&A&16.8&44.45&1.7&37.05&39.72&17.93&51.15&17.59&13.98&15.79&56.6&50.95\\
37&N&N&A&A&2.15&3.7&55.05&39.1&16.68&15.4&20.74&25.88&95.08&51.67&72.9&51.94\\
38&A&N&A&A&28.05&2.3&31.75&37.9&43.13&43.1&10.73&10.83&49.62&48.74&51.81&46.96\\
39&RP&N&A&A&39.1&1.2&32.2&27.5&48.87&17.09&12.74&13.98&53.73&49.69&48.84&46.42\\
40&N&A&A&A&1.85&31.95&32.8&33.4&10.22&10.01&51.36&49.28&47.59&45.2&45.57&42.03\\
41&A&A&A&A&23.82&24.07&23.3&28.82&28&33.36&27.09&34.06&28.2&33.31&30.33&32.08\\
42&RP&A&A&A&32.45&19.85&23.25&24.45&31.96&19.51&30.2&36.61&30.88&37.07&30.62&31.96\\
43&N&RP&A&A&2.05&29.4&35.1&33.45&11.29&11.74&40.95&26.79&52.1&48.22&49.9&46.13\\
44&A&RP&A&A&19.5&30.75&23.25&26.5&30.74&37.33&30.87&18.19&32.55&37.39&34.9&36.44\\
45&RP&RP&A&A&27.98&25.73&21.65&24.63&33.92&17.69&32.4&20.8&33.84&40.51&37.16&38.33\\
46&N&N&RP&A&3.1&8.83&30.98&57.08&17.55&16.86&22.61&27.52&55.04&36.55&83.24&57.16\\
47&A&N&RP&A&24.85&2.45&31.75&40.95&42.59&43.99&12.03&12.92&43.21&29.95&61.02&52.78\\
48&RP&N&RP&A&45.3&2.15&18.25&34.3&53.6&18.56&13.07&15.04&42.84&26.21&56.64&50.47\\
49&N&A&RP&A&3.15&35.55&23.9&37.4&11.76&13.85&55.17&50.8&38.27&24.36&51.92&47.79\\
50&A&A&RP&A&21.3&20.4&29.1&29.2&29.84&34.86&29&34.51&29.68&19.16&36.03&38.58\\
51&RP&A&RP&A&29.35&16.9&30.75&23&32.43&19.45&29.29&34.9&32.44&19.37&34.52&37.1\\
52&N&RP&RP&A&2.7&32.9&25.9&38.5&12.55&13.31&40.54&19.37&39.66&25.41&57.57&50.12\\
53&A&RP&RP&A&17.88&29.38&25.13&27.6&30.45&35.27&32.86&17.42&31.71&20.23&40.88&42.24\\
54&RP&RP&RP&A&26.5&23.15&23.9&26.45&35.53&17.92&33.34&17.22&33.97&20.47&39.83&41.22\\
55&N&N&N&RP&9.3&10.5&16.25&63.95&39.99&35.34&34&33.9&41.14&39.13&79.92&23.62\\
56&A&N&N&RP&51.3&3.35&6.25&39.1&82.98&57.86&18.11&17.88&20.89&25.36&64.8&40.16\\
57&RP&N&N&RP&60.2&3.85&7.4&28.55&62.41&17.32&22.88&23.57&22.62&25.93&52.94&19.07\\
58&N&A&N&RP&2.45&50.45&3.2&43.9&20.84&22.97&79.86&56.9&18.33&20.05&63.81&18.87\\
59&A&A&N&RP&38.3&24.05&1.85&35.8&57.74&52.87&43.6&45.62&12.71&11.92&46.07&29.17\\
60&RP&A&N&RP&25.5&32.55&3.65&38.3&40.35&26.79&50.6&50.02&14.85&15.32&42.77&20.85\\
61&N&RP&N&RP&3.9&47.45&3.35&45.3&23.5&25.54&68.54&18.79&22.41&22.11&68.82&19.52\\
62&A&RP&N&RP&28.2&43.25&2&26.55&52.96&50.66&52.64&18.16&14.42&16.31&48.88&29.95\\
63&RP&RP&N&RP&18.8&56.9&2.95&21.35&42.81&17.92&57.97&17.08&15.72&17.83&44.89&18.82\\
64&N&N&A&RP&3.55&3.45&59.35&33.65&20.06&22.18&19.99&22.9&86.02&55.06&58.01&17.68\\
65&A&N&A&RP&25.8&2.7&38.05&33.45&45.92&47.07&12.89&13.45&52.78&49.93&42.74&25.79\\
66&RP&N&A&RP&49&1.25&32.9&16.85&53.93&17.02&12.98&13.58&50.88&48.23&40.63&16.34\\
67&N&A&A&RP&1.25&31&23.15&44.6&11.56&13.45&52.77&50.25&44.23&44.14&49.75&16.45\\
68&A&A&A&RP&20.1&20.65&20.55&38.7&31.32&36.74&28.95&35.28&28.78&33.68&34.41&19.89\\
69&RP&A&A&RP&27&21.7&19.35&31.95&32.81&20.78&33.95&39.1&29.26&32.51&34.1&16.67\\
70&N&RP&A&RP&2.35&21.8&30.2&45.65&13.04&15.32&43.36&26.24&51.41&48.6&54.12&18.2\\
71&A&RP&A&RP&21.43&28.13&20.6&29.83&34.05&39.57&32.9&18.64&31.68&35.29&35.66&20.1\\72&RP&RP&A&RP&29&22.7&19.3&29&35.61&18.56&34.72&21.02&33.16&39.45&36.87&16.57\\
73&N&N&RP&RP&3.45&8&27.55&61&21.03&22.89&23.04&27.77&50.71&19.12&62.92&17.96\\
74&A&N&RP&RP&37&2.65&30.25&30.1&54.99&51.28&13.65&14.84&39.99&19.49&43.66&28.23\\
75&RP&N&RP&RP&57.3&2.7&20.05&19.95&57.85&16.69&15.79&18.05&42.93&19.19&43.23&16.66\\
76&N&A&RP&RP&2.35&33.95&15.15&48.55&13.87&17.69&53.21&50.43&38.75&17.06&53.6&16.5\\
77&A&A&RP&RP&21.9&19.85&27.6&30.65&34.74&40.23&30.82&37.79&32.72&17.55&35.4&21.15\\
78&RP&A&RP&RP&23.8&23.3&24.5&28.4&35.2&22.5&37.16&43.14&33.87&16.98&36.34&17.68\\
79&N&RP&RP&RP&2.2&22.8&18.1&56.9&14.61&17.61&44.35&18.7&41.36&18.15&58.35&16.97\\
80&A&RP&RP&RP&21.85&26.45&24.1&27.6&37.53&42.78&35.12&17.43&33.7&17.7&37.51&21.3\\
81&RP&RP&RP&RP&24.2&27.43&22.73&25.63&36.84&18.04&37.89&18.32&35.81&17.64&37.89&18.21\\

\hline
\label{ludo_4p_12turns}
\end{longtable}

\begin{figure}[H]
    \centering
    \includegraphics[width=0.6\textwidth]{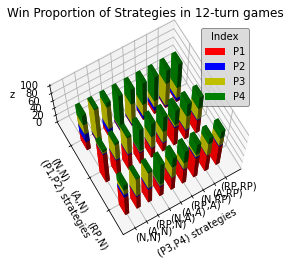} 
    \caption{Win Proportion of strategy combinations 1-27 in 12-turn game} 
    \label{fig:ludo_12turns-1-27} 
\end{figure}

\begin{figure}[H]
    \centering
    \includegraphics[width=0.6\textwidth]{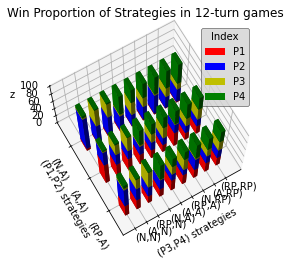} 
    \caption{Win Proportion of first strategy combinations 28-54 in 12-turn game} 
    \label{fig:ludo_12turns-28-54} 
\end{figure}

\begin{figure}[H]
    \centering
    \includegraphics[width=0.6\textwidth]{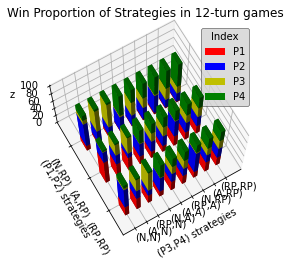} 
    \caption{Win Proportion of strategy combinations 55-81 in 12-turn game} 
    \label{fig:ludo_12turns-55-81} 
\end{figure}

\begin{center}
 {\bf 8-turn, 10-move game}  
\end{center}

\begin{longtable}{|p{0.4cm}|p{0.55cm}|p{0.55cm}|p{0.55cm}|p{0.55cm}|p{0.75cm}|p{0.75cm}|p{0.75cm}|p{0.75cm}|p{0.9cm}|p{0.8cm}|p{0.9cm}|p{0.8cm}|p{0.9cm}|p{0.8cm}|p{0.9cm}|p{0.8cm}|}
\hline
Sl. No. & \multicolumn{4}{c|}{Strategies} & \multicolumn{4}{c|}{Win Percentages} & \multicolumn{2}{c|}{P1 Points} & \multicolumn{2}{c|}{P2 Points} & \multicolumn{2}{c|}{P3 Points} & \multicolumn{2}{c|}{P4 Points} \\ 
\hline
 & P1 & P2 & P3 & P4 & P1 & P2 & P3 & P4 & Mean & Sd & Mean & Sd & Mean & Sd & Mean & Sd \\
\hline
\endhead
\hline
\endfoot

1&N&N&N&N&24.73&24.87&24.82&25.58&38.23&21.81&38.01&22.17&37.86&20.34&39.54&22.01\\
2&A&N&N&N&81.6&6.4&5.3&6.7&80.64&42.32&22.32&12.99&22.44&17.37&25.37&19.41\\
3&RP&N&N&N&78.1&3.5&7.9&10.5&52.55&13.43&27.84&13.57&23.15&18.29&24.62&19.62\\
4&N&A&N&N&6.45&82.95&5.9&4.7&25.11&20.74&76.54&41.14&23.5&13.62&22.98&17.36\\
5&A&A&N&N&53.35&35.2&6.55&4.9&53.5&40.5&38.76&32.21&15.99&10.86&15.62&11.89\\
6&RP&A&N&N&36.45&49.5&7.75&6.3&36.46&17.49&45.85&38.41&18.08&11.78&16.08&13.23\\
7&N&RP&N&N&9.6&77.3&4.1&9&20.79&18.62&51.77&12.88&28.3&13.79&25.8&18.39\\
8&A&RP&N&N&52.35&40.6&3.45&3.6&54.09&43.2&40.26&12.78&17.67&12.18&16.54&13.09\\
9&RP&RP&N&N&30.95&59.95&3.25&5.85&35.95&16.38&44.22&11.67&20.82&13.23&17.89&15.57\\
10&N&N&A&N&5&6.45&80.55&8&20.94&18.53&25.43&20.2&76.84&42.13&25.06&12.94\\
11&A&N&A&N&44.7&6&42.8&6.5&48.19&40.13&13.25&11.23&45.85&36.42&16.13&12.59\\
12&RP&N&A&N&52.05&2.22&41.2&4.53&43.21&14.15&16.58&14.04&46.61&38.71&16.07&11.9\\
13&N&A&A&N&4.4&49&40.4&6.2&13.6&12.88&49.54&37.81&39.96&30.09&16.44&10.65\\
14&A&A&A&N&33.98&31.18&29.05&5.78&31.88&32&27.92&24.1&26.39&21.94&12.29&9.39\\
15&RP&A&A&N&36.08&31.2&27.08&5.63&29.91&14.76&29.43&25.22&26.88&23.54&12.84&9.95\\
16&N&RP&A&N&5.6&31.8&52.45&10.15&13.79&13.38&34.23&16.18&46.99&37.44&19.62&12.06\\
17&A&RP&A&N&34.15&28.3&31.6&5.95&33.19&33.73&26.82&12.63&29.48&25.1&13.16&9.7\\
18&RP&RP&A&N&36.15&28.25&30.35&5.25&30.72&15.14&28.13&14.49&30.48&27.05&13.49&10.1\\
19&N&N&RP&N&6.8&11.05&77.35&4.8&21.26&17.34&22.29&20.74&51.91&13.56&29.19&13.18\\
20&A&N&RP&N&41.75&2.5&52.3&3.45&46.57&37.98&13.14&11.01&43.49&14.71&18.28&13.24\\
21&RP&N&RP&N&45.85&2.2&50&1.95&47.81&13.1&16.74&13.67&47.72&14.33&17.64&13.12\\
22&N&A&RP&N&3.3&45.95&46.65&4.1&14.68&13.6&46.3&37.85&40.4&13.3&19.54&13.66\\
23&A&A&RP&N&26.98&24.68&44.9&3.43&30.88&31.58&26.3&25.17&34.39&13.38&13.59&10.97\\
24&RP&A&RP&N&29.7&29.7&38.15&2.45&31.9&15.55&31.38&29.87&36.39&13.07&13.4&10.6\\
25&N&RP&RP&N&6.2&26.35&63&4.45&15.21&15.39&33.88&15.51&44.72&12.14&21.76&13.92\\
26&A&RP&RP&N&28.7&21.9&47.85&1.55&34.21&36.41&28.29&14.69&37.64&13.26&13.92&11.14\\
27&RP&RP&RP&N&29.5&21.7&45.2&3.6&33.66&15.46&29.64&14.88&39.92&12.06&15.77&12.8\\
28&N&N&N&A&4.7&5.35&5.7&84.25&20.16&13.84&21.07&17.78&23.41&19.29&78.71&41.83\\
29&A&N&N&A&31.5&4.2&4.25&60.05&38.74&34.34&13.32&10.65&14.82&13.6&55.32&38.72\\
30&RP&N&N&A&42.7&3.2&3.45&50.65&40.14&13.2&16.63&13.22&14.96&14.17&50.01&39.04\\
31&N&A&N&A&5.95&41.8&6.25&46&13.44&12.45&46.95&38.81&14.65&12.78&45.53&34.68\\
32&A&A&N&A&33.45&24.65&4.15&37.75&27.25&24.27&24.77&24.56&10.58&9.3&31.69&28.02\\
33&RP&A&N&A&28.33&28.33&6.4&36.93&27.05&13.94&29.61&30.14&11.67&10.54&32.57&31.1\\
34&N&RP&N&A&2.7&53.13&1.73&42.43&12.43&11.18&44.44&14.72&17.16&14.09&49.35&39.23\\
35&A&RP&N&A&21.15&44.15&2.5&32.2&26.4&26.23&33.77&13.26&11.76&11.06&34.03&31.87\\
36&RP&RP&N&A&21.93&47.43&2.65&27.98&28.81&14.28&37.8&13.9&12.7&11.12&31.87&30.15\\
37&N&N&A&A&3.9&3.6&49.6&42.9&13.35&10.93&12.74&12.5&49.29&37.98&40.26&28.95\\
38&A&N&A&A&25.75&4.4&29.25&40.6&24.18&22.29&9.73&8.86&28.66&28.9&32.6&25.47\\
39&RP&N&A&A&43.6&2.5&24.65&29.25&34.23&14.26&10.71&9.7&28.22&26.67&29.35&23.55\\
40&N&A&A&A&2.6&32.88&27.98&36.53&9.81&8.97&32.1&31.83&26.82&24.14&28.44&21.41\\
41&A&A&A&A&23.23&21.38&23.25&32.13&19.09&18.9&17.66&18.17&18.22&17.8&22.48&18.71\\
42&RP&A&A&A&31.8&22.4&20.6&25.2&24.07&13.43&20.59&21.66&19.47&19.69&21.62&18.4\\
43&N&RP&A&A&3.4&32.15&27.95&36.5&10.5&10.04&28.24&14.79&28.82&25.73&29.98&24.26\\
44&A&RP&A&A&18.03&34.23&22.5&25.23&18.55&19.49&23.76&13.53&21.62&22.89&21.96&18.78\\
45&RP&RP&A&A&27.6&27&19.95&25.45&25.05&14.6&24.62&13.46&20.8&20.99&23.02&19.52\\
46&N&N&RP&A&6.35&6.85&31.2&55.6&15.07&11.68&14.48&13.95&33.04&16.33&46.01&34.21\\
47&A&N&RP&A&21.6&3.65&32.4&42.35&26.5&25.76&9.78&8.81&28.68&15.41&35.92&28.43\\
48&RP&N&RP&A&47.47&1.25&20.12&31.17&37.32&13.95&10.66&10.36&29.12&15.02&33.63&29.19\\
49&N&A&RP&A&5.1&30.8&27.2&36.9&10.35&9.35&30.08&30.41&26.39&13.74&31.57&24.78\\
50&A&A&RP&A&22.35&18.73&29.43&29.48&19.27&18.88&18.43&19.63&22.62&12.67&23.12&20.41\\
51&RP&A&RP&A&29.53&19.8&28.18&22.48&24.71&13.27&20.9&22.19&23.69&13.21&21.55&18.37\\
52&N&RP&RP&A&3.58&31.88&25.05&39.48&11.28&10.01&27.79&15.66&27.6&13.54&35.54&28.99\\
53&A&RP&RP&A&19.4&27.25&26.45&26.9&20.61&22.54&23.17&13.98&24.29&13.75&24.93&22.29\\
54&RP&RP&RP&A&28.2&22.7&21.55&27.55&26.77&14.14&23.84&14.08&24.1&13.41&26.49&24.06\\
55&N&N&N&RP&3.2&6.4&10.75&79.65&26.78&15.11&21.07&18.35&21.67&20.66&53.81&13.18\\
56&A&N&N&RP&44.72&5.72&6.8&42.77&43.63&38.11&15.6&10.99&13.66&12.37&37.73&18.23\\
57&RP&N&N&RP&58.95&3.5&4.25&33.3&44.21&11.84&18.7&13.42&15.11&13.39&36.56&15.6\\
58&N&A&N&RP&3.35&37.35&2.85&56.45&16.13&14.16&46.66&41.17&13.58&11.52&44.53&13.78\\
59&A&A&N&RP&32.6&26.5&3.6&37.3&31.81&30.73&27.86&26.89&11.03&9.14&32.28&15.84\\
60&RP&A&N&RP&29.7&31.05&2.95&36.3&28.2&14.63&32.18&30.84&11.36&9.24&30.85&14.55\\
61&N&RP&N&RP&2.2&43.9&2.3&51.6&15.61&14.49&47.28&14.12&16.56&14.55&48.55&13.34\\
62&A&RP&N&RP&21.37&45.47&3.1&30.07&28.33&28.49&37.08&13.25&11.8&11.42&32.95&15.38\\
63&RP&RP&N&RP&21.18&49.63&3&26.18&30.29&14.74&39.72&12.69&13.29&12.21&32.61&14.17\\
64&N&N&A&RP&2.25&3.85&46.9&47&16.87&12.74&14.31&13.34&46.13&39.47&41.07&13.08\\
65&A&N&A&RP&24.28&4.93&30.43&40.35&26.41&25.63&11.6&9.76&29.22&29.34&30.16&14.11\\
66&RP&N&A&RP&46.35&2.9&24.45&26.3&37.15&13.46&12.87&11.36&28.76&29.1&30.22&13.84\\
67&N&A&A&RP&1.9&24.05&23.65&50.4&10.58&10.39&29.53&30.55&26.92&25.52&36.24&13.11\\
68&A&A&A&RP&22.33&16.53&20.85&40.28&21.1&22.91&17.84&17.41&18.2&17.28&26.64&12.8\\
69&RP&A&A&RP&29.48&18.88&17.93&33.7&24.77&13.95&20.58&21.46&19.44&20.46&26.09&13.48\\
70&N&RP&A&RP&1.6&22.6&23.25&52.55&11.2&11.22&29.38&15.8&29.47&29.06&39.07&12.98\\
71&A&RP&A&RP&19.7&29.3&18.25&32.75&21.47&22.37&24.83&13.18&19.69&18.84&27.02&12.62\\
72&RP&RP&A&RP&26.65&25.65&18&29.7&25.97&14.47&25.59&13.27&20.93&20.63&27.88&13.04\\
73&N&N&RP&RP&2.65&4&30.7&62.65&17.98&13.95&14.61&14.45&35.03&15.91&44.2&11.5\\
74&A&N&RP&RP&27.8&2.7&32.45&37.05&30&29.89&10.85&9.14&28.24&15.57&31.29&14.61\\
75&RP&N&RP&RP&53.3&1.85&21.75&23.1&40.66&12.71&12.67&11.27&30.91&15.65&31.61&13.61\\
76&N&A&RP&RP&2&24.5&17.75&55.75&12.22&12.19&29.43&29.57&26.56&14.13&38.68&12.11\\
77&A&A&RP&RP&20&18&29.45&32.55&20.72&21.76&19.05&19.86&23.85&13.09&27.03&13.66\\
78&RP&A&RP&RP&25.5&18.7&24.55&31.25&24.92&14.1&21.85&22.28&24.39&13.41&27.61&13.66\\
79&N&RP&RP&RP&1.6&23.65&22.45&52.3&12.11&11.92&30.8&15.12&29.67&14.26&41.11&11.78\\
80&A&RP&RP&RP&18.4&28.4&23.85&29.35&22.27&22.45&26.46&13.73&24.55&14.22&28.04&13.08\\
81&RP&RP&RP&RP&28.28&23.73&21.9&26.08&27.88&13.89&26.64&14.44&26.3&13.67&28.61&13.56\\

\hline
\label{ludo_4p_8turns}
\end{longtable}

\begin{figure}[H]
    \centering
    \includegraphics[width=0.6\textwidth]{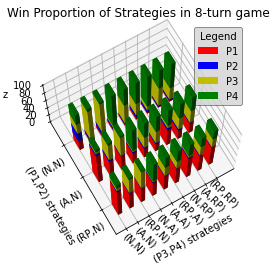} 
    \caption{Win Proportion of strategy combinations 1-27 in 8-turn game} 
    \label{fig:ludo_8turns-1-27} 
\end{figure}

\begin{figure}[H]
    \centering
    \includegraphics[width=0.6\textwidth]{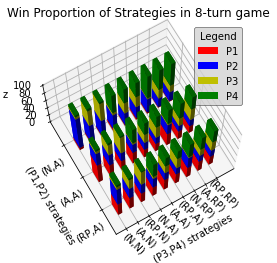} 
    \caption{Win Proportion of first strategy combinations 28-54 in 8-turn game} 
    \label{fig:ludo_8turns-28-54} 
\end{figure}

\begin{figure}[H]
    \centering
    \includegraphics[width=0.6\textwidth]{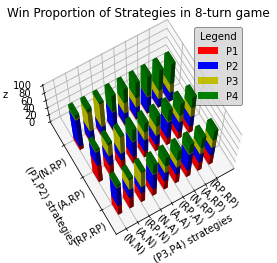} 
    \caption{Win Proportion of strategy combinations 55-81 in 8-turn game} 
    \label{fig:ludo_8turns_55-81} 
\end{figure}

\section{Empirical Nash Equilibrium}
\label{nash}

From the simulated empirical data, Nash equilibrium of strategy combinations is computed. 
Since empirical data is used, an approximate version of Nash equilibrium ($\epsilon$-NE) is computed using 3 different choice of $\epsilon$, i.e, reduction in win percentage by margin less than the specified $\epsilon$ is not considered a significant reduction.

\begin{table}[H]
\centering
\begin{tabular}{|c|c|c|c|}
\hline
\textbf{Turns $\downarrow$ \; $\epsilon$} $\rightarrow$ & \textbf{$\epsilon = 0$} (exact NE) & \textbf{$\epsilon = s.e. = 0.5$} & \textbf{$\epsilon = 2 s.e. = 1$} \\
\hline
\textbf{16 turns} & (RP, RP) & (RP, A), (RP, RP) & (RP, A), (RP,RP) \\
\hline
\textbf{20 turns} & (A, A) & (A, A) & (A, A) \\
\hline
\textbf{24 turns} & (A, A) & (A, A) & (A, A)  \\
\hline
\end{tabular}
\caption{Empirical Nash Equilibria for 2 player, 3-dice version of ludo}
\label{tab:nash_2p}
\end{table}

\begin{table}[H]
\centering
\begin{tabularx}{\textwidth}{|c|X|X|X|}
\hline
\textbf{Turns $\downarrow$ \; $\epsilon$} $\rightarrow$ & \textbf{$\epsilon = 0$} (exact NE) & \textbf{$\epsilon = s.e. = \frac{\sqrt{10}}{2}$} & \textbf{$\epsilon = 2 s.e. = \sqrt{10}$} \\
\hline
\textbf{16 turns} & (RP, RP, RP, A), (RP, A, RP, RP) & (RP, RP, RP, A), (RP, RP, A, RP), (RP, A, RP, RP) & (RP, A, A, RP), (RP, A, RP, A), (RP, A, RP, RP), (A, RP, RP, A), (RP, RP, A, A), (RP, RP, A, RP), (RP, RP, RP, A), (RP, RP, RP, RP) \\
\hline
\textbf{12 turns} & (RP, RP, RP, A) & (RP, RP, RP, A), (RP, RP, RP, RP) & (RP, RP, RP, A), (A, RP, RP, RP), 
 (RP, RP, RP, RP)\\
\hline
\textbf{8 turns} & (RP, RP, RP, A) & (RP, RP, RP, A), (RP, RP, RP, RP)  & (RP, RP, RP, A), (RP, RP, RP, RP)  \\
\hline
\end{tabularx}
\caption{Empirical Nash Equilibria for 4 player, 5-dice version of ludo}
\label{tab:nash_4p}
\end{table}

\subsection{Comparative Discussion}

{\bf 2-player, 3-dice version:}
Three sets of empirical Nash Equilibrium (with varying choices of $\epsilon$) have been computed for three different lengths of the game. The 16-turn game is the only one having two different strategies as Nash equilibrium strategies, namely, (RP,A) and (RP,RP). This reflects the benefits of safe play in the shorter formats, which is also supported by the fact that (RP,RP) is the only exact Nash equilibrium. However, for slightly relaxed $\epsilon$ criteria, it is also beneficial for the second player to play aggressively, possibly due to the second mover's advantage.
\footnote{Sonas (2002) documents a similar 4\% (3\%) advantage for White in classical (Rapid) chess games, where the first mover can influence the selection of opening strategy for the Black by choosing a particular opening.}

The 20-turn and 24-turn games clearly favour aggression, with (A,A) being the only Nash equilibrium strategy for all considered $\epsilon$. Thus, the longer format is clearly more advantageous for the aggressive player. \\
Also, none of the several Nash equilibrium strategy include the naive player, which is intuitive, given the naive player's simple algorithm, which enhances the importance of intelligently devised strategies in this game. \\

{\bf 4-player, 5-dice version:} 
Similar to the 2-player, 3-dice version, three sets of Nash equilibrium have been computed here for three different lengths of the game. 

The 8-turn and 12-turn games have similar Nash equilibrium strategies, with comparatively lesser number of optimal strategies compared to the 16-turn counterpart. (RP,RP,RP,A) is the only exact NE strategy, illustrating the importance of safe play, while showing that aggression is also a preferable option, though only for the $4^{th}$ player. This can be explained by the fact that the crucial $8^{th}$ turn, when the tokens are expected to start entering each other's territory, starts with the $4^{th}$ player rolling the die and getting five die to choose from and potentially make a capture. \footnote{Berthelemy (2023) documents the presence of such a 'tipping move', a move which most likely determines the outcome of a game in chess, a well-established game of skill.}

As the $\epsilon$ criteria is slightly relaxed, (RP,RP,RP,RP) is also considered a Nash equilibrium strategy, highlighting the dominance of safe play in the shorter formats. Interestingly, for the maximal $\epsilon$ considered (= 2 s.e), (A,RP,RP,RP) is a Nash equilibrium strategy for the 12-turn game(but not for the 8-turn). This can be explained similarly as the last point, since the $1^{st}$ player gets a choice of four die in the $8^{th}$ turn, second only to the $4^{th}$ player. Also, the inclusion of two strategies involving an aggressive player as Nash equilibrium in the 12-turn game hints that the aggressive strategy is more preferred in longer formats than shorter ones. 

The 16-turn game has two exact Nash equilibrium strategies, (RP,RP,RP,A) and (RP,A,RP,RP). Quite counter-intuitively, (A,RP,RP,RP) is not a NE strategy, which can be possibly explained by the dominant $4^{th}$ player in this strategy (28.75\% win rate), which has higher probability of capturing the first player following aggression. 
\footnote{Duersch et. al. (2020) documents that if players with similar strength (the $1^{st}$ and $4^{th}$ aggressive and responsible players, respectively), are paired together, the net difference between good players is low, causing the weaker players to perform better.}

(RP,RP,A,RP) is the other Nash equilibrium strategy for $\epsilon = s.e$, showing the improvement of the aggressive strategy with increased game length. 

There are as many as 8 NE strategies for $\epsilon = 2 s.e$. Notably, most of these strategies involve 1 or 2 aggressive players, with the exception of the combination having all responsible players. (RP, A, RP, A) , (A, RP, RP, A) and (RP, RP, A, A), which are all possible combinations of 2 aggressive players with the $4^{th}$ player as aggressive, are Nash equilibrium strategies, again indicating at the $4^{th}$ player's advantage. (RP,A,A,RP) is the other NE strategy, where the two responsible players, each having an advantage in terms of more available rolls at the crucial $8^{th}$ turn, possibly end up capturing and hindering each other's progress. 

\begin{remark}
\label{rem1}
For any fixed $\epsilon$ criteria, the number of Nash equilibrium strategies increases with an increase in the number of turns, indicating a possible decrease in qualitative differences between aggressive and safe play as game length increases.
\end{remark}
\begin{remark}
\label{rem2}
No strategies involving more than 2 aggressive players are considered Nash equilibrium, possibly due to increased interaction between players in these cases, which lets one player improve by deviating to safe play. 
\end{remark}
\begin{remark}
\label{rem3}
None of the game lengths and $\epsilon$ criterion allow any strategy involving the naive player as Nash equilibrium, reflecting the innate importance of well designed skill based strategies in this version of ludo. 
\footnote{This is in agreement with the findings of Rudd \& Miles (2019), who state that players have the ability to change the outcome in their favor only in games of skill.}
\end{remark}
\newpage
\section*{References}

\begin{enumerate}
    \item Diganta Mukherjee \& Subhamoy Maitra (2023), Unveiling the potential and scope of the Online Skill Gaming Industry: Study with technology students and professionals, report submitted to E-Gaming Federation.
    \item Tomašev, N., Paquet, U., Hassabis, D., \& Kramnik, V. (2020). Assessing Game Balance with AlphaZero: Exploring Alternative Rule Sets in Chess. arXiv (Cornell University). \\ https://doi.org/10.48550/arxiv.2009.04374
    \item Prajit K. Dutta (1999), Strategies and Games: Theory and Practice, The MIT Press.
    \item Diganta Mukherjee, Subhamoy Maitra, and Swagatam Das (2023), Role of Skill in the Game of Online Rummy: A Statistical Analysis, March 2023, DOI:10.13140/RG.2.2.12826.11203
    \item Jeff Sonas (2002), The Sonas Rating Formula — Better Than Elo?. ChessBase.com. 22 October 2002.
    \item Barthelemy, M. (2023). Statistical analysis of chess games: space control and tipping points. arXiv (Cornell University). https://doi.org/10.48550/arxiv.2304.11425
    \item Duersch, P., Lambrecht, M., \& Oechssler, J. (2020). Measuring skill and chance in games. European Economic Review, 127, 103472. https://doi.org/10.1016/j.euroecorev.2020.103472
    \item Rudd, D., \& Mills, R. (2019). CAN YOU WIN: GAMES OF SKILL. ResearchGate.
\end{enumerate}

\end{document}